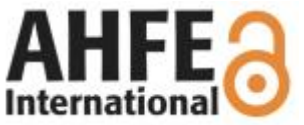

# Does AI Assistance Leave a Temporal Fingerprint? Detecting Overreliance in AI-Assisted Writing and Programming

**Eduardo Davalos[1] and Yike Zhang[2]**

[1]Trinity University, San Antonio, TX 78212, USA
[2]St. Mary's University, San Antonio, TX 78228, USA

## ABSTRACT

The rapid adoption of generative AI has made final artifacts unreliable evidence of student learning, and AI detectors that examine only the finished product are inaccurate and ethically contentious. Process data offers an alternative, but prior work covers only English essay writing. We ask whether AI assistance carries a temporal signature, whether it generalizes from writing to programming, and whether it distinguishes ordinary collaboration from wholesale delegation. We analyze three public corpora: CoAuthor (1,447 keystroke-level co-writing sessions), RealHumanEval (editor telemetry from 243 programmer records), and a pre-LLM CS1 corpus (5.1 million keystrokes) as a human-only baseline, comparing minimal-AI work, collaborative AI use, and simulated wholesale delegation. Three findings emerge. First, the signature generalizes: AI contributions arrive in bursts far outside the author's own baseline in both mediums (paired $d_z$ = 1.13 and 3.54). Second, engagement diverges by medium: 93% of AI-inserted characters survived to writers' final documents, while only 14% of accepted code suggestions survived intact. Third, classifiers using only observable temporal features separate simulated delegation from authentic work nearly perfectly (F1 ≥ 0.997; at most 0.5% of real work misclassified), while ordinary collaboration remains hard to distinguish from unassisted work. Temporal evidence flags wholesale delegation rather than assistance, positioning process visibility as a candidate evidentiary basis for academic integrity, pending validation in authentic coursework.

## INTRODUCTION

When students can produce polished essays and code with conversational agents, the finished artifact no longer reveals how learning occurred. Institutional responses have leaned on AI text detectors, which operate on the final product alone, show unreliable accuracy, and produce contested accusations that erode trust (Deep et al., 2025). The Learning Visibility Framework reframes this as a measurement problem: what is lost when AI enters the assessment loop is visibility into the learning process, and what assessment requires is process-based evidence alongside outcomes (Davalos & Zhang, 2026).

Process data is a promising basis for such evidence: keystroke features distinguish authentic composition from transcription of AI-generated text with high accuracy in English essays (Crossley et al., 2024; Deane et al., 2026; Kundu et al., 2024; Roh et al., 2025). But this evidence covers a single medium. No peer-reviewed study has tested whether AI-inserted code is detectable from temporal process features, and none has compared signatures across mediums. If process evidence is domain-specific, every discipline needs its own instrumentation and validation; if it generalizes, process visibility becomes a scalable foundation for academic integrity across the curriculum.

A second gap concerns what should be detected. The relevant integrity boundary is not whether AI was used, since AI use is often permitted and productive, but whether the student engaged with the material or delegated the work wholesale. Overreliance is a broad construct; we operationalize and test its most delegation-like form, wholesale pasting of AI output followed by cosmetic edits, and are explicit in regard to about the forms this operationalization cannot capture. Whether such delegation harms learning is not tested here; we treat that link as motivated by prior work on cognitive offloading (Risko & Gilbert, 2016).

To our knowledge, this paper contributes: (1) the first temporal-signature analysis of AI-inserted content in programming process data; (2) the first cross-domain comparison of such signatures between writing and programming; (3) a provenance-exact temporal reanalysis of CoAuthor, a combination left open between prior studies (Yang et al., 2026; Zeng et al., 2024); and (4) the first three-condition reliance spectrum, operationalized via a grounded simulation and classified from observable temporal features, showing that delegation, not assistance, is what temporal evidence detects well.

## RELATED WORK

Writing-process research shows text emerging in bursts terminated by pauses (P-bursts) or revisions (R-bursts), with burst length indexing translation capacity (Chenoweth & Hayes, 2001); pause thresholds near two seconds conventionally separate transcription from higher-order planning (Wengelin, 2006), and keystroke logging methodology is well established (Leijten & Van Waes, 2013).

Building on this, Crossley et al. (2024) distinguished authentic from transcribed essays with 99% accuracy: transcription showed long linear bursts, few revisions, and short pauses. Related results appear in large-scale assessment, biometric, and non-English settings (Deane et al., 2026; Kundu et al., 2024; Roh et al., 2025), all in writing. In programming, AI-code detection has remained artifact-based, operating on submitted code rather than its production (Ramachandra et al., 2026), while pre-LLM process research targeted compile-event metrics, identification, and plagiarism deterrence (Ihantola et al., 2015; Leinonen et al., 2016; Edwards & Hart, 2023). Emerging systems match assistant responses to subsequent edits (He et al., 2026) or annotate reliance from code-edit logs (Jin et al., 2026); blind detection of AI-inserted code from temporal features remains untested. Scalable watermarking now marks model output itself (Dathathri et al., 2024), yet a watermark shows only that a model was involved, not how: delegation with light edits and genuine collaboration carry the same mark.

Two adjacent lines deserve contrast. Overreliance has been linked to interaction-log patterns via unsupervised clustering among AI users (Liu et al., 2026), and AI assistance has been detected from temporal behavior in abstract search tasks (King et al., 2025); neither analyzes writing or programming process telemetry, and neither trains a supervised classifier across reliance conditions. Cross-domain comparisons of AI use to date rely on conversation content, not process telemetry.

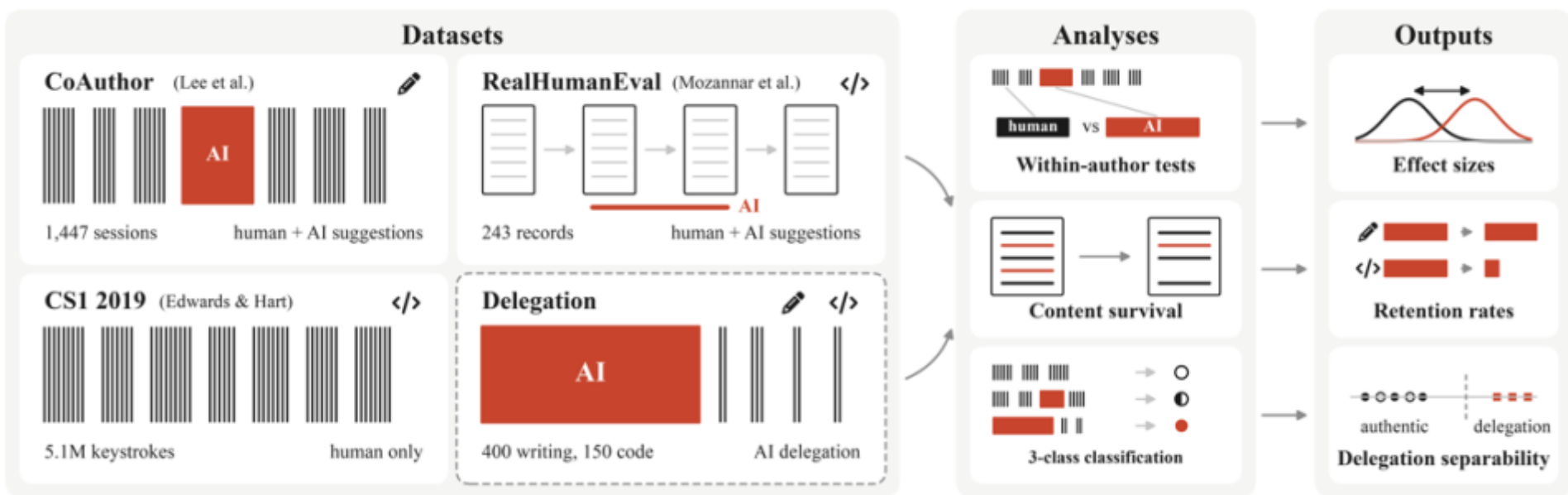

**Figure 1: Study overview**. Left: the four datasets, with red marking AI-origin content: tagged keystroke streams (CoAuthor), snapshots with AI acceptances (RealHumanEval), human-only keystrokes (CS1), and simulated wholesale delegation (dashed). These logs feed three analyses (within-author comparisons, content survival, three-class classification) yielding effect sizes, retention rates, and delegation separability.

## DATA

We analyze three public, de-identified corpora and one simulated condition, which means no new human data were collected. Neither AI-era corpus involves students in authentic coursework: CoAuthor writers were crowdworkers, and RealHumanEval participants completed short lab-style tasks. A study overview is shown in Fig. 1.

**CoAuthor (writing).** 1,447 sessions by 63 writers composing stories and essays with GPT-3 suggestions (Lee et al., 2022). Every timestamped insertion is tagged user or API; replaying sessions through a character-level provenance buffer yields exact per-origin insertion, deletion, and survival accounting. Sessions with fewer than three AI insertions form a **minimal-AI** condition (n = 194; 56 contain none; writers self-selected into declining suggestions); the rest are collaborative (n = 1,253).

**RealHumanEval (programming).** Editor telemetry (Mozannar et al., 2025): 243 participant records, snapshots at 15 to 45 s cadence, 5,204 autocomplete events (589 accepted), and 1,055 chat interactions including copied text. Diffing consecutive within-task snapshots yields insertion and deletion deltas; intervals containing an acceptance or chat copy are classed AI-containing. 219 traces met length criteria: human-only (zero AI events, n = 88) and collaborative (two or more, n = 131); 45 participants support the paired analyses.

**CS1 keystrokes (external baseline).** The 2019 CS1 dataset (Edwards & Hart, 2023): 5.1 million keystrokes from 487 students in a pre-LLM semester; every event is necessarily human.

**Simulated delegation.** No public dataset captures wholesale pasting of AI output with light edits, so we simulate it (400 writing sessions; 150 programming traces), labeled as such throughout. Writing sessions paste one to three blocks (90 to 98% of the document), then read through and edit 3 to 12% of characters at a real writer's resampled pace with reading-scale pauses; programming traces place one or two large deltas early, then edits and gaps resampled from human-only traces. The simulation asks whether observable features separate this behavior from real work.

## METHOD

Provenance-labeled analyses compare AI-inserted spans against the author's own production: burst segmentation at the standard 2 s pause threshold (Wengelin, 2006), insertion sizes, inter-keystroke intervals (IKI), deletion rates by origin, and survival into the final artifact. In writing, character provenance yields exact survival; in code, we measure graded survival as the fraction of an AI span's 20-character normalized shingles present in the task's final snapshot, over spans of at least 40 characters. The two measures are qualitatively comparable, not numerically equivalent.

Spectrum analyses use only observable, label-free session features. Writing (eight): largest, 95th-percentile, and median burst; burst coefficient of variation (CV); median IKI; pause rate; deletion ratio; product-process ratio (Crossley et al., 2024). Programming (seven): largest, 95th-percentile, and median snapshot delta; delta CV; largest-delta share of all insertions; deletion ratio; mean snapshot gap. IKI, pause, and product-process features are writing-only because snapshots preclude them. Signatures are evaluated against each author's own baseline, following evidence that individual typing distributions are stable (Leinonen et al., 2016).

Paired comparisons use Wilcoxon signed-rank tests with Cohen's $d_z$ (mean of paired differences over their standard deviation, on log-transformed measures; not comparable to between-groups d). Real-condition contrasts use Mann-Whitney tests, Bonferroni-corrected per domain; contrasts with the simulated condition are descriptive (medians, Cliff's delta), since inference against a researcher-specified generator is not meaningful. Classification uses random forests (300 trees, balanced class weights) under stratified 5-fold cross-validation, reporting per-class F1 and confusion structure. Fold-internal median imputation reproduces all results; CoAuthor releases no writer identifiers, so folds are session-level. All analyses run from public data with a fixed seed.

## RESULTS

### The insertion-size signature generalizes across mediums

AI contributions arrive in chunks the same author's typing rarely produces. In writing, AI insertions averaged 67.8 characters against the writer's own 35.7-character bursts (geometric means; $d_z$ = 1.13, $p$ < .001, n = 1,253), yet exceeded the writer's 95th-percentile burst in only 19% of sessions: the distributions separate but overlap (Fig. 2a). The contrast holds at 0.5, 1, 2, and 3 s pause thresholds ($d_z$ = 3.73, 2.27, 1.13, 0.52). In programming the effect is stronger at event granularity: AI-containing snapshot intervals inserted 188 characters against 51 for human-only intervals ($d_z$ = 3.54, $p$ < .001, n = 45; Fig. 2b). Writers also paused a median 2.8 s immediately after accepting a suggestion, nineteen times their 144 ms typing interval.

The pre-LLM CS1 corpus anchors how anomalous such arrivals are: across 1,331 traces (median IKI 176 ms), the median burst was 6 characters, the 95th percentile 38, and only 0.2% of traces contained any single insertion above 100 characters. Insertions of the size AI routinely produces are, practically, absent from human-only keystroke records.

## The afterlife of AI content diverges by medium

Writers keep AI text; programmers rework or discard it. Only 7.2% of AI-inserted characters were ever deleted against 9.6% of self-authored ones ($d_z$ = -0.20, $p$ < .001), so 92.8% of AI prose survived to the final document. Programmers showed the opposite: 13.7% of accepted autocomplete spans survived intact, 62.2% were discarded, and chat-copied code was never fully retained (81.1% discarded; n = 452 autocomplete and 238 chat spans).

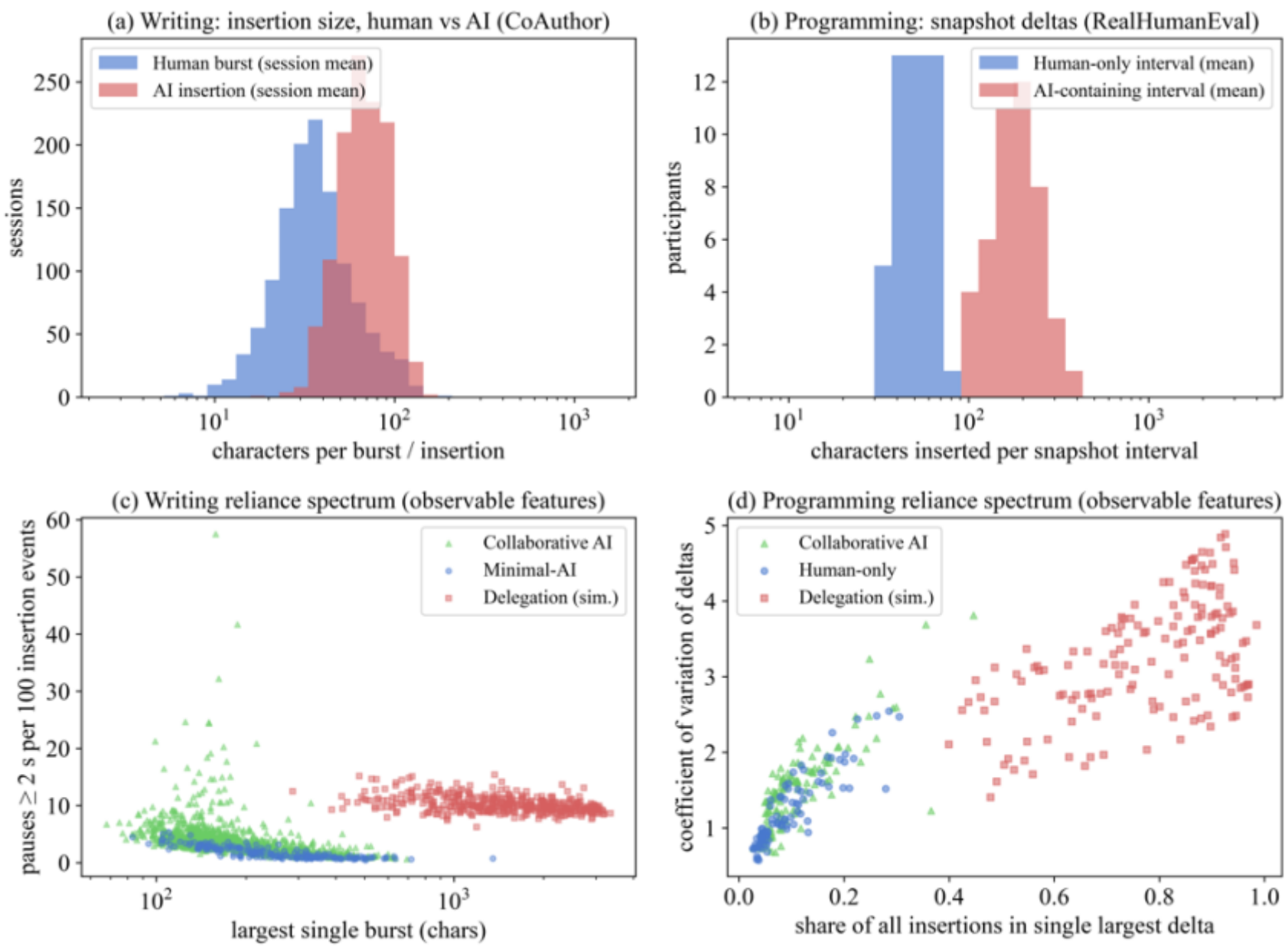

**Figure 2**: **Temporal signatures and the reliance spectrum**. (a, b) Mean insertion sizes (per session in writing, per participant in programming) for human versus AI contributions. (c, d) Sessions positioned by observable features; simulated delegation (squares) separates from both real conditions (circles: baseline; triangles: collaborative), which overlap.

Figure 3 summarizes the contrast. One candidate explanation is verification pressure: code faces executable feedback that forces engagement, prose does not; but the comparison confounds medium with genre, population, suggestion length, and logging granularity, and high retention of short, actively chosen suggestions after an evaluation pause is also consistent with appropriate acceptance. We therefore present cognitive offloading (Risko & Gilbert, 2016) as one permitted interpretation, on which writing is the higher-risk medium for silent delegation. These results agree with short-horizon vendor telemetry (Ziegler et al., 2024), field-scale removal of AI completions in 31% of 53.6K developer trajectories (Liang et al., 2026), and topic-level persistence of AI content in essays (Bhat et al., 2026); our measure is complete-span survival in the final artifact.

## Temporal evidence detects delegation, not assistance

Wholesale delegation stands out immediately in coarse logs; ordinary AI use does not. Figure 4 shows why: real sessions accumulate text steadily, with AI acceptances as small steps in a writer-paced trajectory, while delegation is a near-vertical jump followed by a long, nearly flat polish phase. Simulated delegation differs categorically on observable features (Table 1; Fig. 2c, d): largest bursts of 1,477 characters against 167 to 242, and 82% of code insertions arriving in one delta against 8%. Between the real conditions, five of eight writing and five of seven code features differ significantly (Bonferroni-corrected, marked * in Table 1); against the simulated condition, Cliff's delta on the dominant feature is +0.98 to +1.00, versus -0.39 and +0.17 between real conditions.

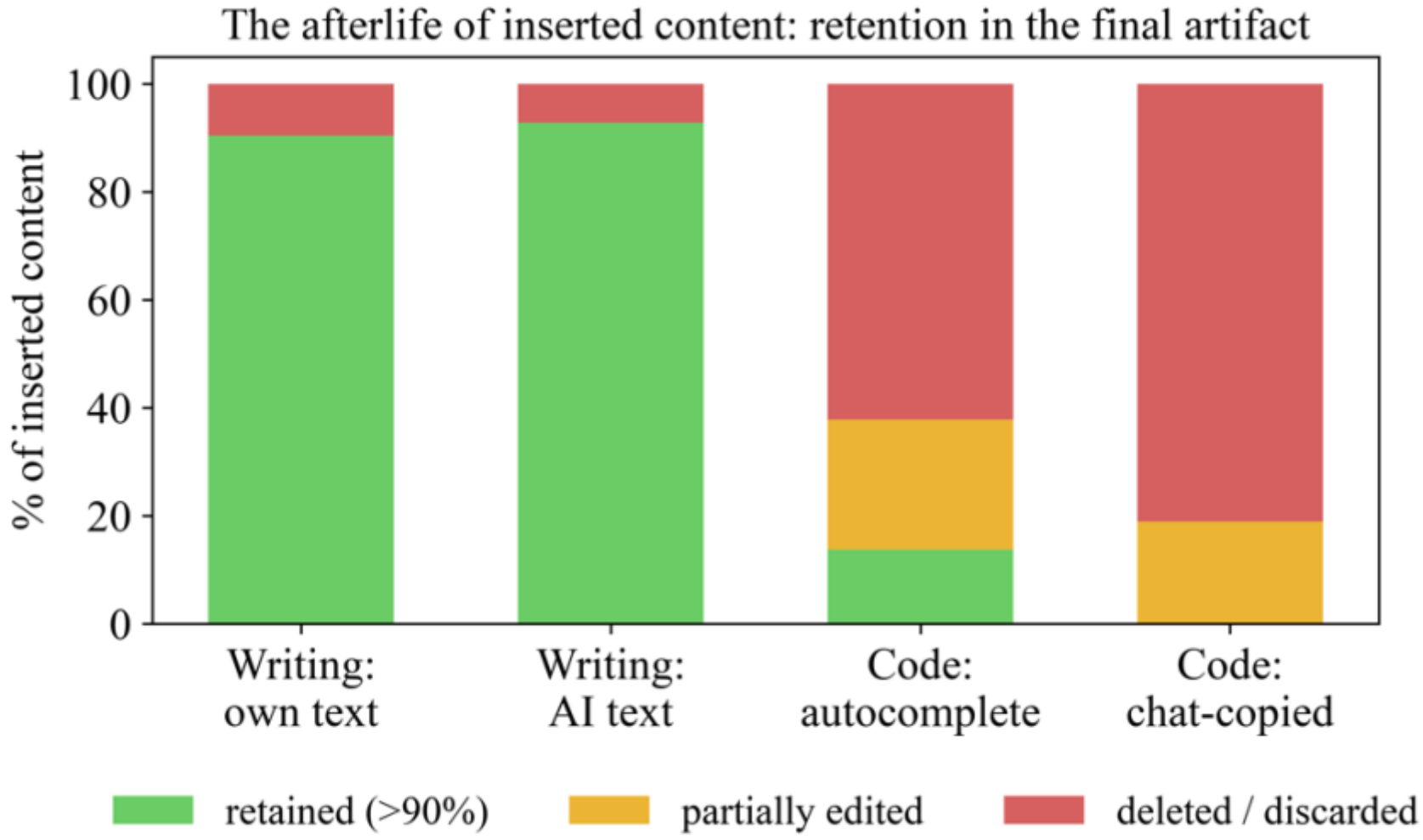

**Figure 3**: **The afterlife of inserted content**. Writing survival uses exact character provenance; code survival uses 20-character shingle matching.

**Table 1.** Condition medians on observable (label-free) features (excerpt).

| **Feature (writing)** | **Minimal-AI** | **Collaborative** | **Delegation (sim.)** |
|---|---|---|---|
| **Largest burst (chars) *** | 242 | 167 | 1,477 |
| **Burst CV** | 0.99 | 0.95 | 2.81 |
| **Pauses ≥ 2 s / 100 events *** | 1.5 | 3.3 | 10.0 |
| **Median IKI (ms) *** | 112 | 144 | 145 |
| **Feature (programming)** | Human-only | Collaborative | Delegation (sim.) |
| **Largest delta (chars) *** | 167 | 354 | 685 |
| **Delta CV *** | 1.08 | 1.47 | 3.19 |
| **Largest-delta share** | 0.08 | 0.08 | 0.82 |
| **Deletion ratio *** | 0.19 | 0.49 | 0.04 |

* Bonferroni-significant between the two real conditions; simulated-condition contrasts are descriptive only.

Random forests on these features separate the three conditions at 91.8% accuracy in writing (macro AUC 0.967) and 79.4% in programming (0.913). The error structure is the finding (Table 2): delegation is detected nearly perfectly in both domains, the two real conditions confuse only each other, and just 1 of 1,447 real writing sessions (0.07%) and 1 of 219 real traces (0.46%) were ever misclassified as delegation. Confusion between minimal-AI and collaborative work carries no integrity implication, since neither represents misconduct.

**Table 2.** Three-condition classification (5-fold CV, observable features only).

| Condition (n writing / n code) | Writing F1 | Programming F1 |
|---|---|---|
| **Minimal-AI / Human-only (194 / 88)** | 0.610 | 0.590 |
| **Collaborative AI (1,253 / 131)** | 0.940 | 0.701 |
| **Delegation, simulated (400 / 150)** | 0.998 | 0.997 |
| **Accuracy / macro AUC** | 0.918 / 0.967 | 0.794 / 0.913 |

Three checks say strong but not trivial. A single-threshold rule (largest burst > 511 characters) catches 97% of simulated delegation but false-flags 1.7% of real sessions; the classifier reaches 0.07% because delegation pairs concentrated production with altered pause and variability structure, a conjunction no real session exhibits. Calibrating the simulator to realistic reading-scale timing changed delegation F1 only from 0.999 to 0.998. And the magnitude matches the published human-data benchmark: Crossley et al. (2024) report 99% accuracy separating real transcription from composition, with the same signature. Further robustness: restricting the writing baseline to the zero-AI subset strengthens the secondary findings (largest-burst gap 322 vs 167; baseline F1 rises to 0.694; delegation unchanged); removing the top feature lowers delegation F1 only to 0.996 and 0.977; and a chunked-paste evasion variant is still caught 82.5% of the time, though paced retyping would evade timing-only features entirely.

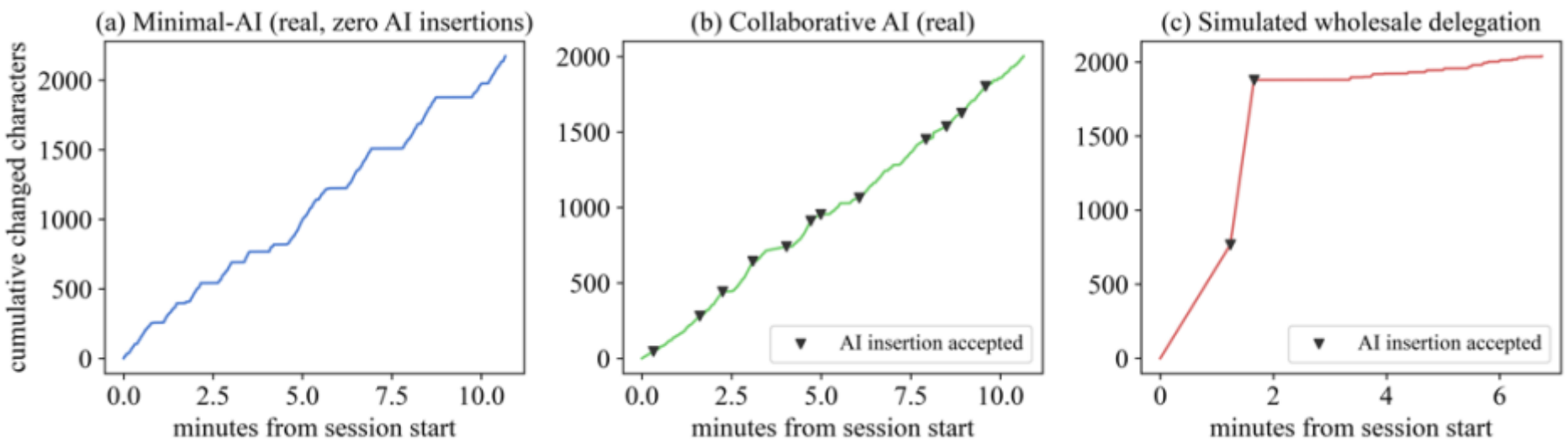

**Figure 4: Example timelines**: cumulative changed characters over time for (a) a real minimal-AI session, (b) a real collaborative session with AI acceptances marked, and (c) a simulated delegation session under the calibrated timing model.

## DISCUSSION

**Process evidence flags delegation, not assistance.** Ordinary AI collaboration is only weakly separable from minimally assisted work, and that is the desirable outcome: students using AI as permitted should not be flagged, and in our results, they essentially never are (at most 0.5% of real work classed as delegation). Wholesale delegation, by contrast, produces a categorical signature that survives without provenance labels or keystroke resolution. The separability is partly by construction, since concentrated production defines the simulated behavior; what the result establishes is sensitivity and observability, not prevalence or field accuracy. The other half of the asymmetry, that collaboration resists detection, is corroborated by a deployed keystroke screening tool that reports missing hybrid

AI use (Asher et al., 2026) and by the modest accuracies of binary keystroke classifiers (Kundu et al., 2024).

**What this method cannot detect, and what a flag can never mean.** Writers retaining 93% of AI prose verbatim is overreliance under uncritical-acceptance definitions (Liu et al., 2026), yet it is temporally indistinguishable from collaboration because it arrives in small, evaluated increments: temporal evidence covers the delegation end of the spectrum only. Conversely, a delegation-shaped timeline has benign look-alikes: drafting in an external editor, dictation, resuming offline work, or pasting instructor templates. For both reasons, and because F1 against a behavior model does not translate into field positive predictive value at realistic base rates, these signals must never function as automated accusation instruments. A flag can open a conversation over a shared timeline; it cannot substantiate an integrity finding.

**The medium moderates engagement.** The afterlife inversion, near-complete character survival of AI prose against minority intact-span survival of AI code, is to our knowledge the first cross-domain measurement of post-acceptance engagement. It suggests assessment risk is not uniform across disciplines: where output faces no executable check, AI content flows into final artifacts with little forced re-engagement; writing environments might import code's verification pressure, for example by prompting reflection on accepted suggestions.

**Implications for educators and tooling.** Every feature used here is computable from coarse, content-free event logs at cadences as slow as 45 s, so any environment that already autosaves could surface reliance indicators without new surveillance machinery. What matters is what happens next: instructor and student walk through the timeline together, and either the concern dissolves or it becomes formative feedback; grading the process portfolio alongside the product is the natural complement. We would hold any deployment to four principles:

1. **Disclose:** students are told in advance, in plain language, what is logged.
2. **Symmetric evidence:** students see the same timeline instructors see.
3. **Conversation:** a flag can open a discussion, not a final verdict.
4. **Audit before use:** false-positive rates are measured across subgroups first, since pause and burst profiles differ across, for example, first- and second-language writers (Roh et al., 2025).

Process visibility that is covert, asymmetric, or unaudited would reproduce the harms of detector-based enforcement it aims to replace (Davalos & Zhang, 2026).

## LIMITATIONS

The delegation condition is simulated: the features detect a behavior we specified, not one we observed in the wild. Still, the specification is empirically grounded and consistent with consented transcription experiments (Crossley et al., 2024; Roh et al., 2025). Real students may not behave like our model: timing can be forged (Condrey, 2026), paced retyping would evade timing-only features, and no outcome data yet link the behavior to learning harm.

The minimal-AI condition is self-selected (the zero-AI analysis addresses only contamination); session-level cross-validation, forced by absent writer identifiers, may inflate the weak minimal-AI versus collaborative separation but cannot manufacture the delegation result; the 2 s pause threshold is a convention (direction holds from 0.5 to 3 s); temporal resolution and survival measures differ across domains, so comparisons rest on effect directions; The dataset CoAuthor's 30-token suggestion cap makes writing estimates conservative; paired programming analyses cover 45 of 243 records; and both corpora record AI arrival as discrete acceptances in short lab tasks, so our claims concern the shape and afterlife of AI contributions, not the occurrence of a paste.

## CONCLUSION

Across 1,447 writing sessions, 219 AI-era programming traces, and 1,331 pre-LLM keystroke traces, AI assistance left a consistent temporal signature: contributions arriving in bursts far beyond the author's baseline, retained nearly verbatim in prose and heavily reworked in code. From observable features alone, wholesale delegation was nearly perfectly separable from authentic work, with under 0.5% of real work ever flagged; ordinary collaboration was not, and did not need to be. Pending validation in authentic coursework, process visibility can supply the evidence integrity decisions require evidence of disengagement rather than tool use, at logging granularities coarse enough to respect privacy.

## DATA AND CODE AVAILABILITY

All corpora are public: CoAuthor (coauthor.stanford.edu), RealHumanEval (Hugging Face, CC0), and the 2019 CS1 keystroke dataset (Harvard Dataverse, CC BY). Analysis code will be made available in a public project repository.

## AI USE DISCLOSURE

Claude (Anthropic) was used for the preparation of this manuscript. The authors inspected and verified the text and take full responsibility for the content.

## REFERENCES

Asher, M.W., Gold, G., Chen, E., Carvalho, P.F. (2026). Chatbots are undermining crowdsourced research in the behavioral sciences: Detecting AI-assisted cheating with a keystroke-based tool. Advances in Methods and Practices in Psychological Science 9(1).

Bhat, A., Aubin Le Quéré, M., Naaman, M., Jakesch, M. (2026). Reactive writers: How co-writing with AI changes how we engage with ideas. In: CHI 2026. arXiv:2603.10374.

Chenoweth, N.A., Hayes, J.R. (2001). Fluency in writing: Generating text in L1 and L2. Written Communication 18(1), 80–98.

Condrey, J. (2026). On the insecurity of keystroke-based AI authorship detection: Timing-forgery attacks against motor-signal verification. arXiv:2601.17280.

Crossley, S., Tian, Y., Choi, J., Holmes, L., Morris, W. (2024). Plagiarism detection using keystroke logs. In: EDM 2024.

Dathathri, S., See, A., Ghaisas, S., Huang, P.-S., McAdam, R., Welbl, J., et al. (2024). Scalable watermarking for identifying large language model outputs. Nature 634, 818–823.

Davalos, E., Zhang, Y. (2026). AI misuse in education is a measurement problem: Toward a learning visibility framework. arXiv:2603.07834.

Deane, P., Zhang, M., Hao, J., Li, C. (2026). Using keystroke dynamics to detect nonoriginal text. Journal of Educational Measurement 63(1), e12431.

Deep, P.D., Edgington, W.D., Ghosh, N., Rahaman, M.S. (2025). Evaluating the effectiveness and ethical implications of AI detection tools in higher education. Information 16(10).

Edwards, J., Hart, K. (2023). Review of CSEDM data and introduction of two public CS1 keystroke datasets. Journal of Educational Data Mining 15(1), 1–31.

He, K., Ma, Q., Chen, V., Chi, W., Wu, T. (2026). RECAP: An end-to-end platform for capturing, replaying, and analyzing AI-assisted programming interactions. In: ACL 2026 System Demonstrations.

Ihantola, P., Vihavainen, A., Ahadi, A., Butler, M., Börstler, J., Edwards, S.H., et al. (2015). Educational data mining and learning analytics in programming: Literature review and case studies. In: ITiCSE-WGR '15, 41–63.

Jin, H., Yoo, M., Han, J., Chen, Z., Ahn, S.-Y., Wang, X. (2026). RelianceScope: An analytical framework for examining students' reliance on generative AI chatbots in problem solving. In: L@S 2026. arXiv:2602.16251.

King, T., Gurney, N., Miller, J.H., Ustun, V. (2025). Detecting AI assistance in abstract complex tasks. In: HCII 2025. arXiv:2507.10761.

Kundu, D., Mehta, A., Kumar, R., Lal, N., Anand, A., Singh, A., et al. (2024). Keystroke dynamics against academic dishonesty in the age of LLMs. In: IEEE IJCB 2024. arXiv:2406.15335.

Lee, M., Liang, P., Yang, Q. (2022). CoAuthor: Designing a human-AI collaborative writing dataset for exploring language model capabilities. In: CHI 2022.

Leijten, M., Van Waes, L. (2013). Keystroke logging in writing research: Using Inputlog to analyze and visualize writing processes. Written Communication 30(3), 358–392.

Leinonen, J., Longi, K., Klami, A., Vihavainen, A. (2016). Automatic inference of programming performance and experience from typing patterns. In: SIGCSE '16.

Liang, J.T., Bairathi, M., Chi, W., Talwalkar, A., Subramani, N., Chen, V. (2026). Learning from 53.6K real-world developer edits of AI-generated code. arXiv:2607.25130.

Liu, C., Zhou, Q., Shen, X., Liu, X.B., Wu, T., Chen, X.A. (2026). Behavioral indicators of overreliance during interaction with conversational language models. In: CHI '26, article 790. ACM. doi:10.1145/3772318.3790332.

Mozannar, H., Chen, V., Alsobay, M., Das, S., Zhao, S., Wei, D., et al. (2025). The RealHumanEval: Evaluating LLMs' abilities to support programmers. Transactions on Machine Learning Research.

Ramachandra, A., Chaudhary, S., Tran, J., Desai, R., Pang, A., Salloum, M. (2026). Detecting AI-generated code in introductory programming courses. In: SIGCSE TS '26, 894–900. ACM.

Risko, E.F., Gilbert, S.J. (2016). Cognitive offloading. Trends in Cognitive Sciences 20(9), 676–688.

Roh, J., Kumar, S., Ngo, T. (2025). LLM-assisted cheating detection in Korean language via keystrokes. In: IEEE IJCB 2025. arXiv:2507.22956.

Wengelin, Å. (2006). Examining pauses in writing: Theory, methods and empirical data. In: Computer Key-Stroke Logging and Writing, 107–130. Elsevier.

Yang, K., Cheng, Y., Zhao, L., Raković, M., Swiecki, Z., Gašević, D., et al. (2026). Ink and algorithm: Exploring temporal dynamics in generative AI-assisted writing. British Journal of Educational Technology.

Zeng, Z., Liu, L., Sha, L., Li, Y., Yang, K., Liu, Z., et al. (2024). Detecting AI-generated sentences in human-AI collaborative hybrid texts. In: IJCAI 2024.

Ziegler, A., Kalliamvakou, E., Li, X.A., Rice, A., Rifkin, D., Simister, S., et al. (2024). Measuring GitHub Copilot's impact on productivity. Communications of the ACM 67(3), 54–63.